\makeatletter
\def\blfootnote{\xdef\@thefnmark{}\@footnotetext}
\makeatother

\documentclass[
  aps,
  prb,
  twoside,
  twocolumn,
  floatfix
  showkeys,
  superscriptaddress,
]{revtex4}

\usepackage{amsmath}
\usepackage{amssymb}
\usepackage{graphicx}
\usepackage{tabularx}
\usepackage{dcolumn}
\usepackage{longtable}
\usepackage{footnote}
\usepackage{appendix}

\begin{document}

\newcommand{\bec}{\begin{center}}
\newcommand{\ec}{\end{center}}
\newcommand{\be}{\begin{equation}}
\newcommand{\ee}{\end{equation}}
\newcommand{\beqn}{\begin{eqnarray}}
\newcommand{\eeqn}{\end{eqnarray}}
\newcommand{\bet}{\begin{table}}
\newcommand{\ent}{\end{table}}
\newcommand{\bib}{\bibitem}

\newcommand{\sect}[1]{Sect.~\ref{#1}}
\newcommand{\fig}[1]{Fig.~\ref{#1}}
\newcommand{\Eq}[1]{Eq.~(\ref{#1})}
\newcommand{\eq}[1]{(\ref{#1})}
\newcommand{\tab}[1]{Table~\ref{#1}}

\renewcommand{\vec}[1]{\ensuremath\boldsymbol{#1}}
\renewcommand{\epsilon}[0]{\varepsilon}

\newcommand{\cmt}[1]{\emph{\color{red}#1}%
  \marginpar{{\color{red}\bfseries $!!$}}}



\title{
Rotation misorientated graphene moir\'{e} superlattices on Cu(111): 
classical molecular dynamics simulations and
scanning tunneling microscopy studies
}


\author{P. S\"ule}
\affiliation{Research Centre for Natural Sciences,
Institute for Technical Physics and Materials Science
Konkoly Thege u. 29-33, Budapest, Hungary}

\author{M. Szendr\H{o}}
\affiliation{Research Centre for Natural Sciences,
Institute for Technical Physics and Materials Science
Konkoly Thege u. 29-33, Budapest, Hungary}

\author{C. Hwang}
\affiliation{Center for Nanometrology, Korea Research Institute
of Standards and Science, Daejeon, Republic of Korea
}

\author{L. Tapaszt\'o}
\affiliation
{Research Centre for Natural Sciences,
Institute for Technical Physics and Materials Science
Konkoly Thege u. 29-33, Budapest, Hungary}


\date{\today}

\begin{abstract}
 Graphene on copper is a system of high technological relevance, as Cu is one of the most widely used substrates for the CVD growth of graphene. However, very little is known about the details of their interaction. One approach to gain such information is studying the superlattices emerging due to the mismatch of the two crystal lattices. However, graphene on copper is a low-corrugated system making both their experimental and theoretical study highly challenging. Here, we report the observation of a new rotational moir\'{e} superlattice of CVD graphene on Cu(111), characterized by a periodicity of ($1.5 \pm 0.05$) nm and corrugation of ($0.15 \pm 0.05$) $\hbox{\AA}$ , as measured by Scanning Tunneling Microscopy (STM). To understand the observed superlattice we have developed a newly parameterized 
Abell-Tersoff potential for the graphene/Cu(111) interface fitted to nonlocal van der Waals density functional theory (DFT) calculations. The interfacial force field with time-lapsed classical molecular dynamics (CMD) provides superlattices in good quantitative agreement with the experimental results, for a misorientation angle of ($10.4 \pm 0.5^{\circ}$), without any further parameter adjustment. Furthermore, the CMD simulations predict the existence of two non-equivalent high-symmetry directions of the moir\'{e} pattern that could also be identified in the experimental STM images.
\\

{\em keywords}: graphene, moire patterns, superlattices, atomistic and nanoscale simulations, molecular dynamics simulations

\end{abstract}

\maketitle



\section{Introduction}

\footnotetext[1]{Corresponding author E-mail: sule@mfa.kfki.hu (P\'eter S\"ule)}
\footnotetext[2]{Corresponding author tel.: +361392222/1909 (P\'eter S\"ule)}

  Chemical vapor deposition (CVD) on transition metal surfaces
represents a promising and scalable approach to achieve
reasonably uniform high-quality monolayer graphene for device applications
\cite{CVD1,CVD2,CVD3}.
  Polycrystalline copper foils have widely been used as substrate for the CVD growth of
graphene \cite{CVD3,Li}. 
On the other hand the Cu(111) facet  
has been proposed as
one of the preferred surfaces for the large-scale uniform high-quality monolayer
graphene growth \cite{Gao}.
 The characterization and interpretation of various graphene (gr) superstructures,
such as the moir\'{e} supercells \cite{Batzill,rev} and the corresponding nanoscale topography 
requires sophisticated experimental and theoretical methods \cite{DFT:Ru-Stradi,DFT:Ru-Hutter,grru13}.
Recent studies show that weakly adhered graphene on various substrates
exhibits multiple oriented rotated moir\'{e} regions with different misorientation angles \cite{Gao,STM-grcu,JACS-grcu,grcu2,grcu:LEEM,rotPt,rotRu,rotIr2,rotPt2,rotIr3,grni,rotIr4}.

 In systems with stronger adhesion (e.g. gr/Ru(0001)) the appearance of rotated moire superstructures is still under debate \cite{grru13,SXRD:25X25,rotRu,Parga}.
In a recent study it has been shown that the moir\'{e} hills (called moirons)
are nonequivalent and therefore the large coincident supercell is the
unit cell of moir\'{e} patterning in gr/Ru(0001) \cite{grru13} confirmed by
other experimental works \cite{SXRD:25X25}.
Even in rigid lattice approximation in coincident supercells
lattice missfit vanishes and this superlattice can be taken as the
unit cell of moir\'{e} patterning \cite{Hermann}.
The graphene overlayer often exhibits an additional
longer range topographic modulation,
besides the apparent periodicity,
when local geometric distortions are considered (buckling, local lateral distortions and/or lattice rotation) \cite{Hermann}.

 It has also been shown that
a few types of rotated moir\'{e} patterns could coexist in gr/Pt(111) and gr/Ir(111).
In gr/Cu(111) the occurrence of rotated gr grains is much less studied \cite{Gao}
as compared to other substrates such as Ir(111), Pt(111), Ni(111),
which can be attributed to the imaging difficulties in this important low-corrugated system with large repeat distance \cite{Gao,JACS-grcu}.
Until now, two types of rotated domains have been identified:
one with nearly vanishing rotation angle $\Theta$ (denoted as R0) and large moir\'{e} supercells
(the cell size is around $6$ nm) and another with $\Theta \approx 7\,^{\circ}$ (R7) 
with a much smaller periodicity of $\sim 2$ nm
\cite{Gao,STM-grcu,JACS-grcu}.
A bump-to-hump corrugation of $\xi \approx 0.35$ $\hbox{\AA}$ has been measured by STM for the aligned gr/Cu(111) \cite{STM-grcu}. For rotation misoriented
samples no corrugation data is available.

 In this paper we 
show that another rotated phase can also exist on Cu(111)
with so far the smallest periodicity of 1.5 nm and the largest rotation angle of 10.4$^{\circ}$ (R10).
Experimental (STM) and theoretical classical molecular dynamics (CMD)
methods have been employed to characterize the new rotated phase.
A new CMD force field is developed for gr/Cu(111) to
analyze the moir\'{e} superstructures in atomic detail
and the height variation is measured by STM along different symmetry directions
of the gr/Cu(111) system.
 It has been shown recently that the development of a new interfacial
force field provides the adequate description of the prototypical gr/Ru(0001)
system \cite{Sule13}.
 First principles calculations (such as DFT) have widely been
used in the last few years to understand the corrugation of nanoscale gr sheets
on various substrates \cite{DFT:Ru-Hutter,DFT:Ru-Stradi,DFT:Ru_Wang,DFT:Ru_corrug},
modelling larger systems, above $1000$ carbon atoms
remains, however challenging, especially when geometry optimization
is included. 
The minimal supercell of the gr/Cu(111) system
includes a few thousands of carbon atoms which definitely
exceeds the size limit of accurate DFT geometry optimizations
and/or {\em ab initio} DFT molecular dynamics.
 Here we show that using a new
DFT adaptively parameterized interfacial Abell-Tersoff potential \cite{Abell,tersoff}
one can quantitatively reproduce even the fine structure off the experimentally observed surface reconstructions of gr on Cu(111) (moir\'{e} superstructures).

\begin{table}[t]
\caption[]
{
The fitted Abell-Tersoff parameters for the graphene/Cu(111) interface.
}
\begin{ruledtabular}
\begin{tabular}{lr}
C-Cu &      \\
\hline%
A (eV)      & 977.7958178 \\
B (eV)      & 320.7794950 \\
$\lambda_1$ & 3.1308174  \\
$\lambda_2$ & 2.0455965      \\
$\gamma$    & 0.0883168 \\
c           & 40.9755961    \\
d           & 0.9528753  \\
h           & 0.9050528    \\
$R_c$ ($\hbox{\AA}$)         & 4.1978605   \\
$D_c$ ($\hbox{\AA}$)         & 0.4794477      \\
$\beta$ ($\hbox{\AA}^{-1}$)  & 1.0    \\
$\lambda_3$                  & 1.5527866 \\
n,m                          & 1   \\
\hline
\end{tabular}
\end{ruledtabular}
\footnotetext[1]{
The parameters have been fitted to small flat gr/Cu(111) systems.
Notations are the same as used in ref. \cite{Sule13} (supplementary material)
and on the web page of lammps \cite{lammps}.
$\lambda_3$ is denoted as $\mu$ and $h=cos \Theta_0$ in the supplementary material of ref. \cite{Sule13}.
}
\label{T1}
\end{table}

\section{Methodology}

\subsection{STM setup}

 Scanning tunneling microscopy (STM) is one of the most suitable characterization tools for studying moir\'{e}-superlattices of graphene emerging due to the mismatch of the atomic lattices of graphene with various substrates. This is mainly due to its high (atomic) resolution capability and the ability to reliably detect
extremely small height variations, provided that the variation of the local density of states
(LDOS) is negligible as compared to topographic features. However, a significant redistribution of the LDOS affecting the corrugation measured by
STM is only expected for high (nanometer radius) local curvatures of graphene membranes
and particularly strong substrate interactions (e.g. graphene on Ru, Ni). STM investigations have been successfully employed to investigate moire-patterns emerging in graphene on Ru (0001) \cite{STM:grru}, Ir(111) \cite{STM:grir}, Rh(111) \cite{STM:grrh} as well as in nonmetallic substrates such as 6H-SiC(0001) \cite{STM:sic} and h-BN \cite{STM:bn}. For the case of graphene on Cu(111), neither the curvature, nor the substrate interaction is strong enough to substantially alter the local electronic density of states distribution; consequently, the measured image is expected to be of topographic origin. This is confirmed by the fact that the measured corrugation was not sensitive to the applied bias voltage. 

 Our  STM measurements have been performed at low temperature (78 K) and under UHV conditions. The investigated graphene samples were grown on polished Cu (111) single crystal surfaces by Chemical Vapor  Deposition at  990 C$^{\circ}$ and 1250 mtorr base pressure, using a methane/hydrogen (20sccm:5sccm) gas mixture as precursor. 
After growth the samples have been transferred into the UHV system. No annealing of the sample was necessary prior to STM measurements.

\subsection{Simulation rules}

 Classical molecular dynamics has been used as implemented
in the LAMMPS code (Large-scale Atomic/Molecular Massively Parallel Simulator) \cite{lammps}.
The graphene  layer has been placed commensurately on the substrate
since the lattice mismatch is small in gr/Cu(111).
However, even this tiny misfit is sufficient 
to form
partly registered positions (alternating hexagonal hollow and ontop sites) which
leads to a moir$\acute{e}$ superstructure.

 Periodic triclinic (rhombic) simulations cells have been constructed with $85 \times 85$
and $255 \times 255$ gr-unit cells.
We find these structures to be suitable for simulations leading to
commensurate superstructures.
The systems are carefully matched at the cell borders in order to 
give rise to perfect periodic superstructures.
Arbitrary system sizes lead to non-perfect matching at the borders.
Moreover,
nonperiodic cells lead to unstable moir\'{e} patterns due to the
undercoordinated atoms at the system border which cannot be handled
by the present force field with CMD.
Nonperiodic structures can be, however, optimized by simple
minimizers which also provide moir\'{e} patterns.
Further refinement of the pattern can be obtained by
time-lapsed CMD.
The moir\'{e} pattern is extremely sensitive to weak effects during CMD such as 
e.g. the improperly treated border atoms and/or the arising tensile stress or strain at the simulation cell border.

 Isobaric-isothermal (NPT ensemble) simulations (with Nose-Hoover thermostat and a prestostat) were carried out at 78 K (STM measurements were also performed at this temperature). Vacuum regions were inserted
between the slab of the gr-substrate system to ensure
the periodic conditions not only in lateral directions (x,y) but also in the direction perpendicular to the gr sheet (z).
The variable time step algorithm has been exploited.
The codes OVITO \cite{ovito} and Gnuplot
has been utilized for displaying atomic and nanoscale structures \cite{Sule13,Sule_2011}.

 The molecular dynamics simulations allow the optimal lateral
positioning of the gr layer 
and the minimization of the lattice misfit.
The relaxation of the systems has been reached in 3 steps:
first conjugate gradient geometry optimization (cg-min) with a subsequent simulation box relaxation (boxrel) of the rhomboid simulation cell has been carried out.
Finally variable time step CMD simulations have been utilized in
few tens of a thousand simulation steps to allow the further reorganization
of the system under thermal and pressure controll (NPT, Nose-Hoover
thermostat, prestostat).
Therefore we use in general the combined cg-min/CMD simulations.
  Time-lapsed CMD (TL-CMD) has been used at 78 K in order to average the morphology over longer timescale and also to account for the effect of temperature. We found that 10000 steps are generally sufficient for a stable moir\'{e}
pattern. The pattern remains stable for time averages of much longer simulations.

 The AIREBO (Adaptive Intermolecular Reactive Empirical Bond Order) 
potential
has been used for the graphene sheet \cite{airebo}.
For the Cu substrate, a recent embedded atomic method (EAM) \cite{EAM:Cu} potential is employed.
For the C-Cu interaction we developed a new Abell-Tersoff-like
angular-dependent potential \cite{tersoff} (this potential goes beyond
the level of pairwise interactions and bond angle dependence has also been
explicitly included).
In the Abell-Tersoff potential file (lammps format) the C-C and Cu-Cu interactions are ignored (nulled out).
The CCuC and CuCCu out-of-plane bond angles were considered only. 
The CuCC and CCuCu angles (with in-plane bonds) are ignored in the applied model.
Considering these angles requires the specific
optimization of angular parameters which leads to the polarization
of angles that does not fit to the original Abell-Tersoff model.
 

\subsection{Parameter fitting}

 We used typical small gr/Cu(111) configurations (with flat graphene sheet)
representatives for binding registries of hollow, top-fcc,  top-hcp
and bridge alignments (see e.g. ref. \cite{Sule13}.). The
potential energy curve (PEC) of the rigid gr-Cu(111) separation 
has been calculated by nonlocal VdW-DFT \cite{LMKLL} using the SIESTA code \cite{SIESTA}.
Then using a code developed by us \cite{potfit} the interfacial Abell-Tersoff potential
has been fitted to these DFT PECs.
The following conditions had to be satisfied by the parameter set:
(i) minimal rhomboid supercell edge size ($d \approx 6.1$ nm) for flat aligned gr
(ii) proper topology of the gr-surface: hump-and-bump morphology with a corrugation of $\xi \approx 0.35 \pm 0.05$ $\hbox{\AA}$.
Hollow-humps (moir\'{e} hills) and ontop-bumps (wells) are required 
as it has been found in other gr/substrate systems (e.g. gr/Ru(0001) \cite{Sule13}).
(iii) interface energy: adsorption or adhesion energy $E_{adh} \approx 0.11 \pm 0.02$ eV/atom
(iv) correct interfacial distances: $d_{C-Cu} \approx 3.1 \pm 0.1$ $\hbox{\AA}$.
Among these requirements we imposed directly only
conditions (iii) and (iv) under parameter fitting (corrugation has
not directly been included in the parameter fitting set).
However, the new parameter set also satisfied automatically conditions (i)-(ii).
Conditions (iii)-(iv) seem to be sufficiently strict to restrict
the parameter space in order to account for the morphology, structure
and energetics of the moir\'{e} superstructures of gr/Cu(111).
The obtained parameter set is shown in Table I.
A more detailed technical discussion
of parameter fitting can be found in Ref.
\cite{Sule2}.

\subsection{{\em Ab initio} DFT calculations}

 First principles DFT calculations have also been carried out for
calculating the adhesion energy per carbon atoms
vs. the C/Cu distance for a small ideal system with a flat graphene layer.
The obtained PECs
can be compared with the similar curve of MD calculations.
We also calculate the DFT potential energy curves of
various binding registries of gr including the hollow and ontop configurations
(atop-fcc and hcp) and also the bridge one.

 For this purpose we used 
the SIESTA code \cite{SIESTA,SIESTA2} which utilizes atomic centered numerical basis set.
The SIESTA code and the implemented Van der Waals functional (denoted as DF2,
LMKLL in the code \cite{SIESTA2}) 
successfully employed in several cases 
for gr (see e.g recent refs. \cite{sies1,sies2}).
We have
used Troullier Martin, norm conserving, relativistic pseudopotentials in fully separable
Kleinman and Bylander form for both carbon and Cu.
A double-$\zeta$ polarization (DZP) basis set was
used.
In particular, 16 valence electrons are considered for Cu atoms
and 4 for C atoms.
Only $\Gamma$ point is used for the k-point grid in the SCF cycle.
The real space grid used to calculate the Hartree, exchange and correlation
contribution to the total energy and Hamiltonian was 300 Ry (Meshcutoff).

\begin{figure*}[hbtp]
\label{F2}
\begin{center}
\includegraphics*[height=6cm,width=8cm,angle=0.]{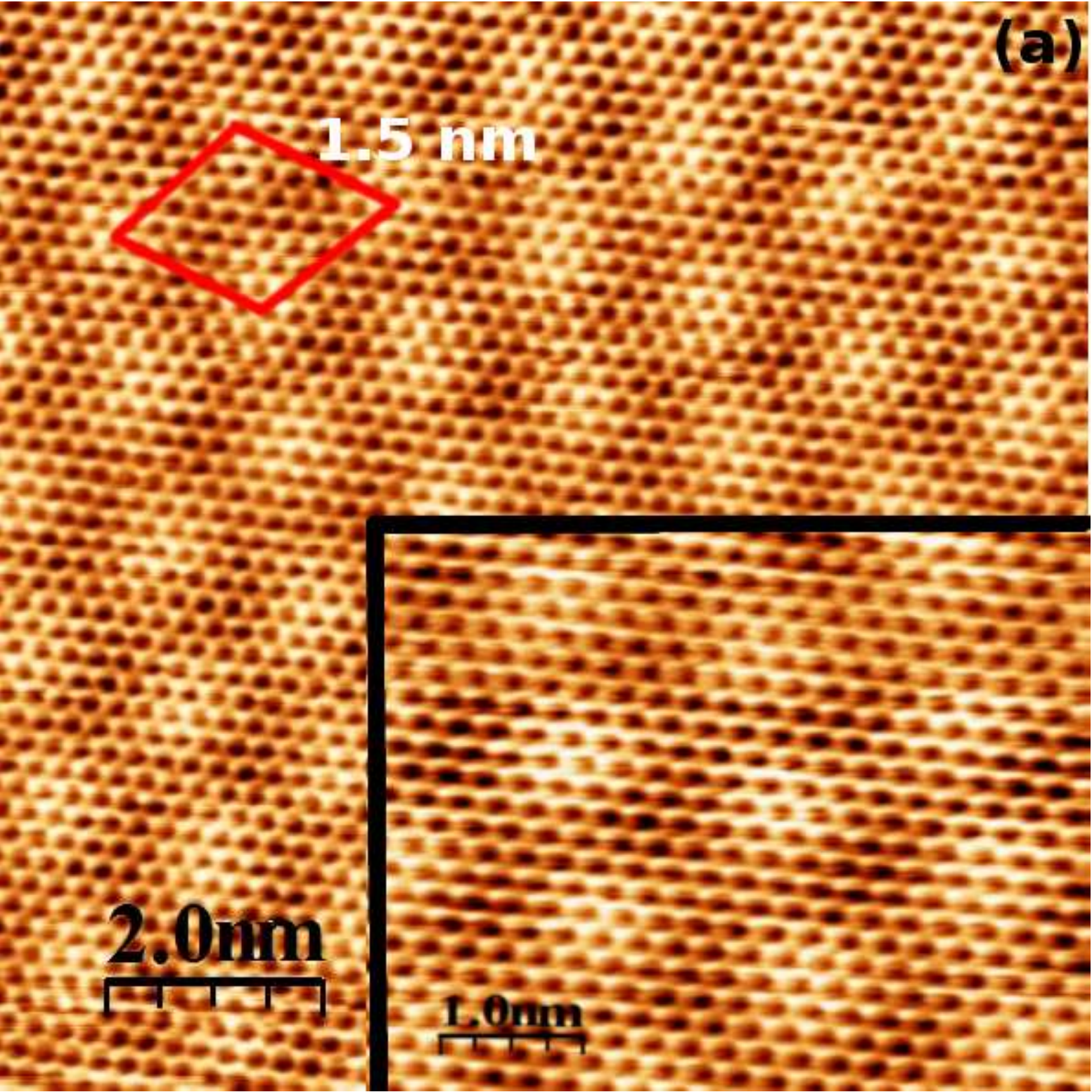}
\includegraphics*[height=6cm,width=8cm,angle=0.]{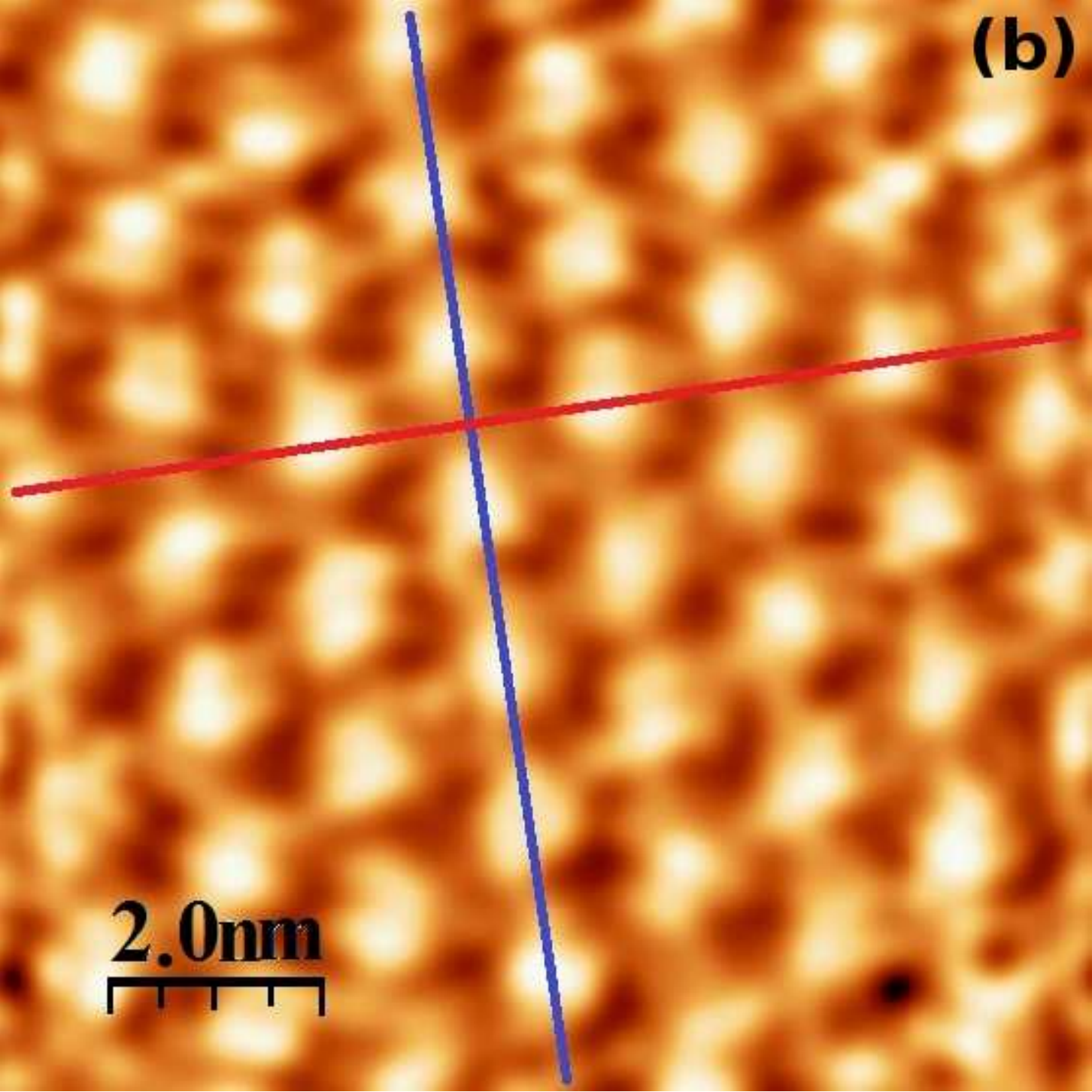}
\includegraphics*[height=6cm,width=8cm,angle=0.]{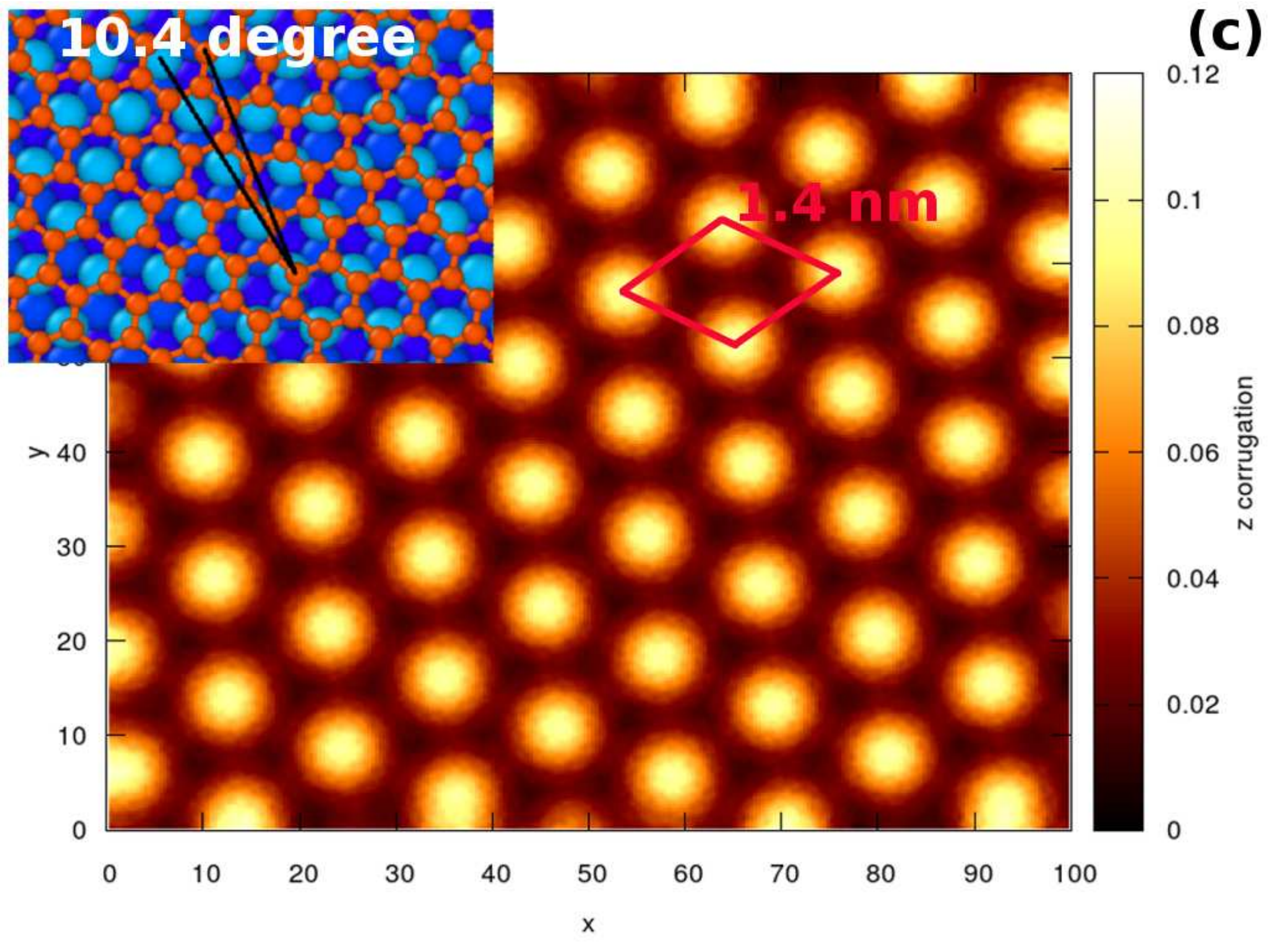}
\includegraphics*[height=6cm,width=8cm,angle=0.]{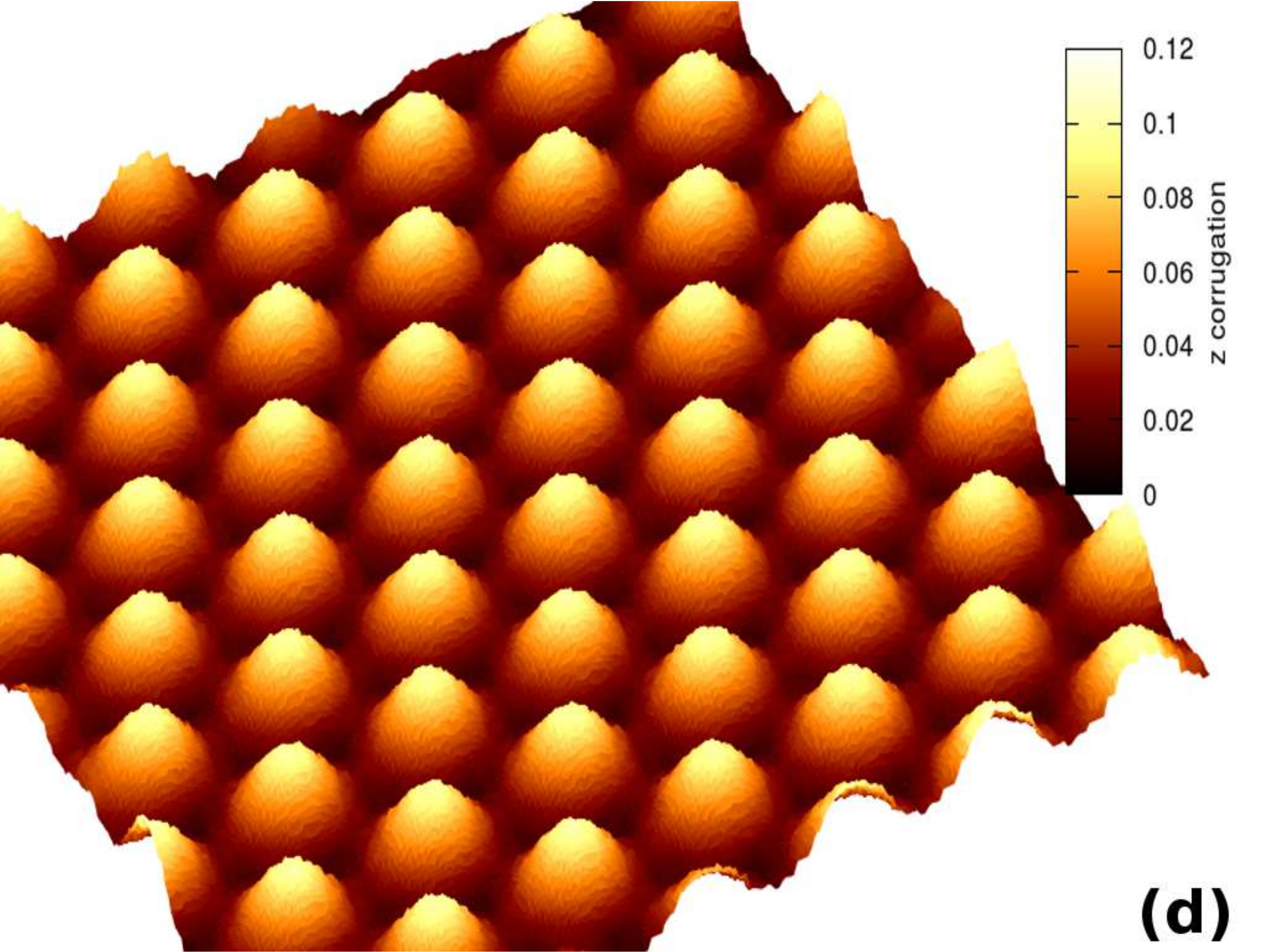}
\caption[]{
 The results of STM measurements and CMD simulations on graphene/Cu(111) moir\'{e} superstructures
at the rotation angle of $\Theta \approx 10.5 \pm 0.5^{\circ}$.
 (a)-(b) Experimental STM images
displaying moir\'{e} superlattice with atomic resolution (a) and when
the atomic corrugation has been removed by Fourier filtering to make the details
of the superlattice more apparent.
(b) The
 directions for measuring the height profiles are also shown
(along the long (red) and short (blue) diagonal of the rhombic supercell).
In the Inset of (a) a magnified image is also shown.
(c) Simulated topographic image. The x,y axes and corrugation are given in
$\hbox{\AA}$.
The minimal ($1 \times 1$) supercells is also shown with a red rhombus.
{\em Inset}: 
The measured rotation angle in the CMD simulations
(the angle between the zig-zag line of carbon atoms 
and the line of (110) Cu atoms).
(d) CMD simulated height profile (3D image) with the moir\'{e} humps (protrusions).
Corrugation is given in $\hbox{\AA}$.
}
\end{center}
\end{figure*}

\begin{figure*}[hbtp]
\label{F3}
\begin{center}
\includegraphics*[height=6cm,width=8cm,angle=0.]{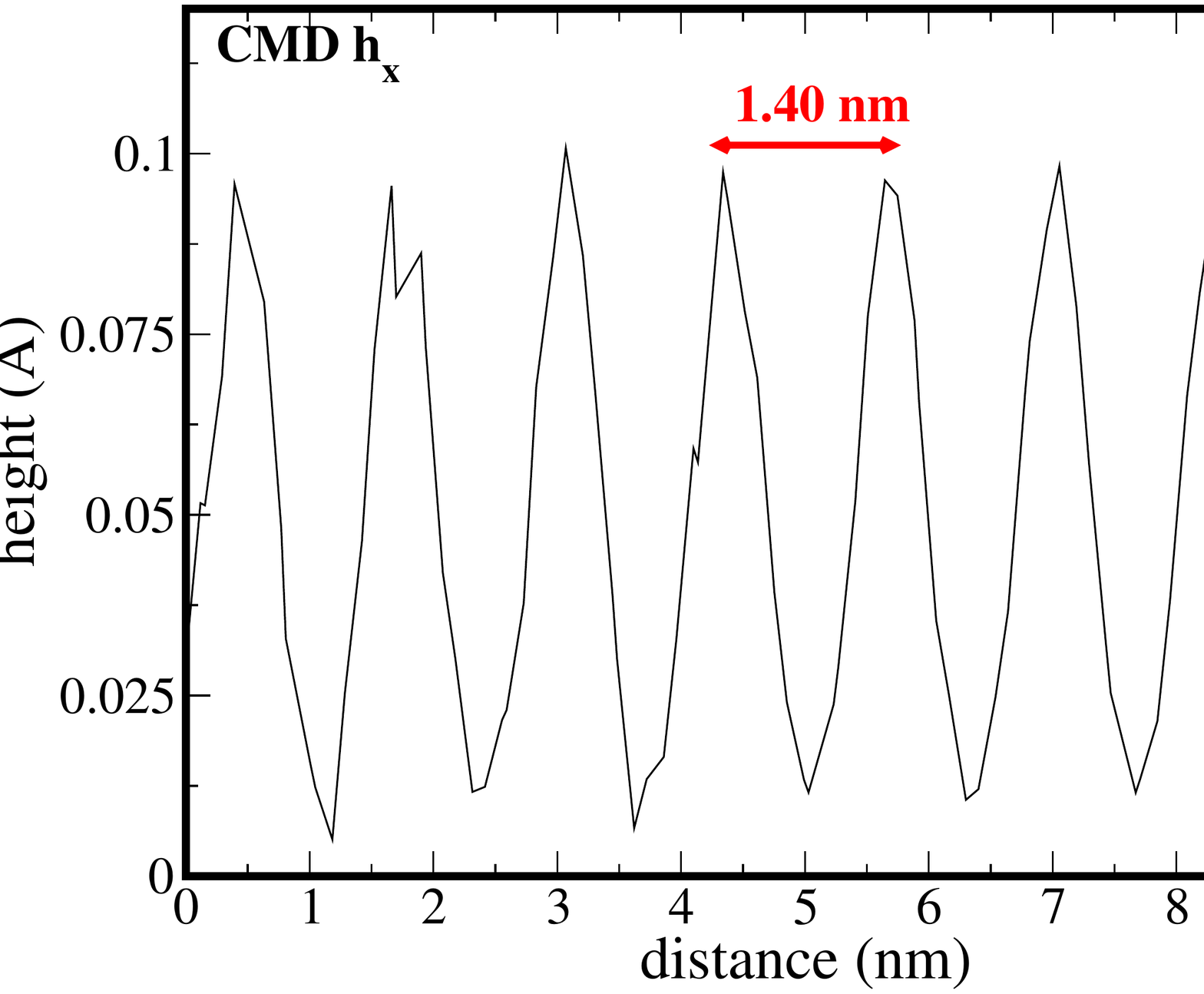}
\includegraphics*[height=6cm,width=8cm,angle=0.]{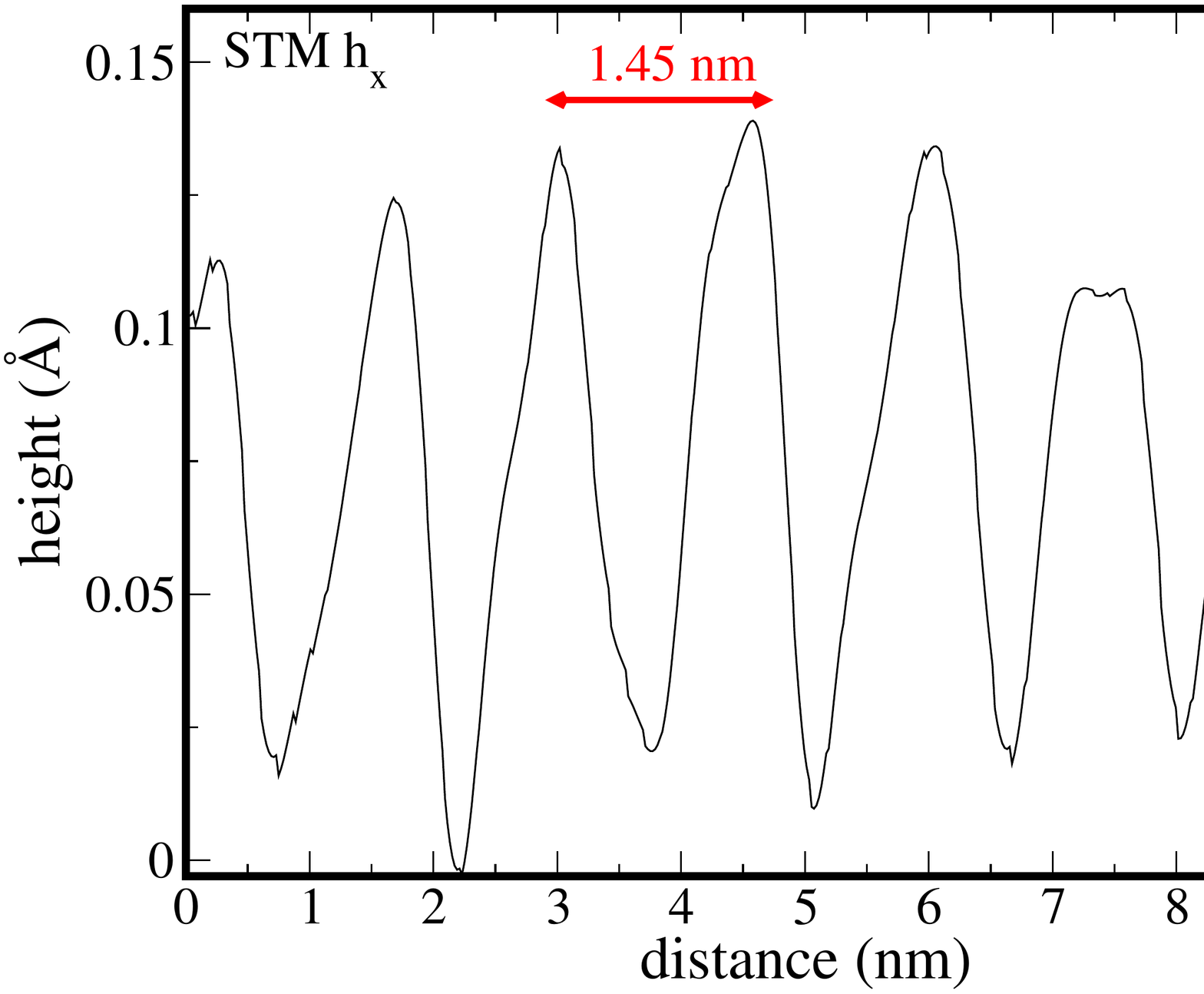}
\includegraphics*[height=6cm,width=8cm,angle=0.]{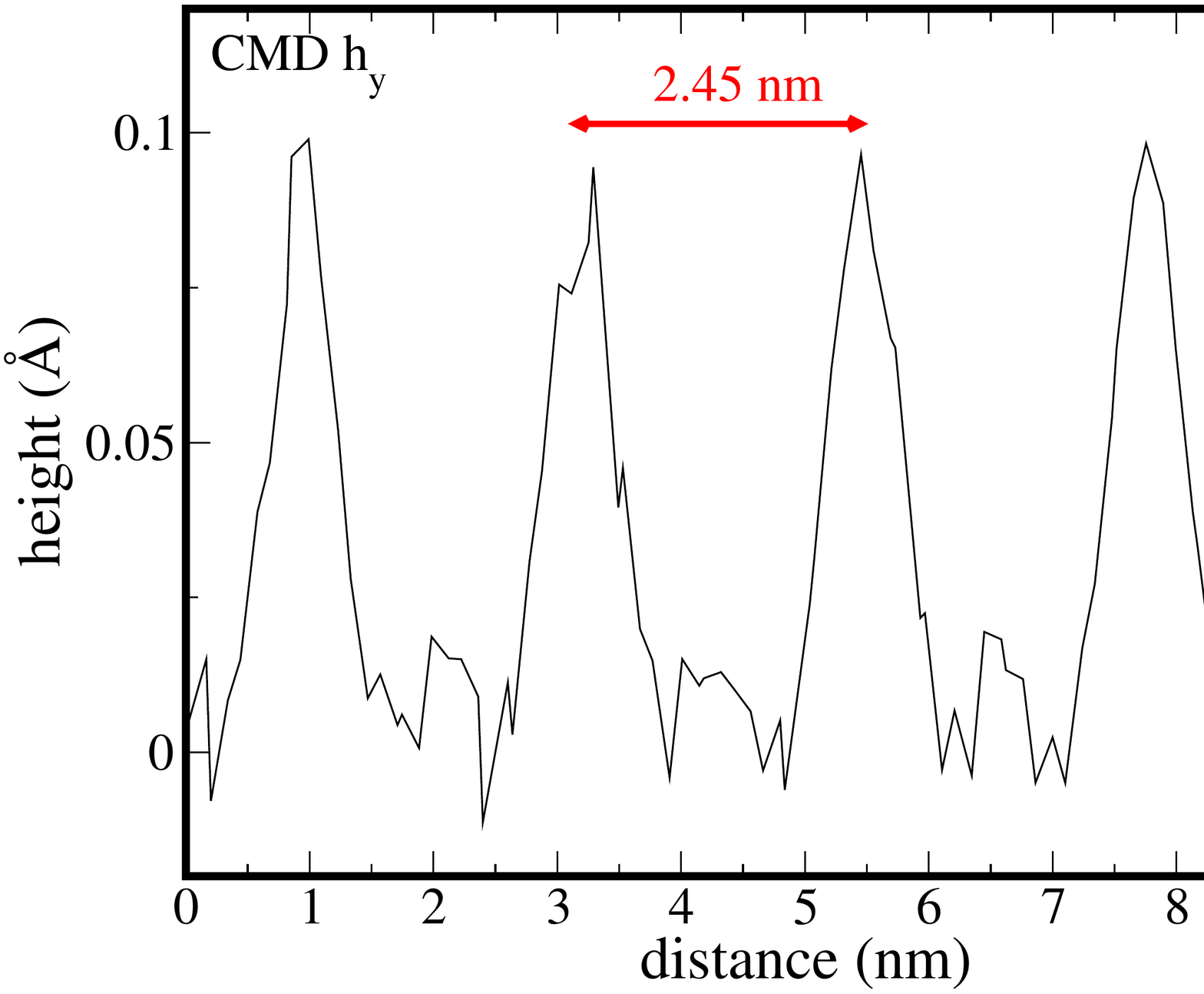}
\includegraphics*[height=6cm,width=8cm,angle=0.]{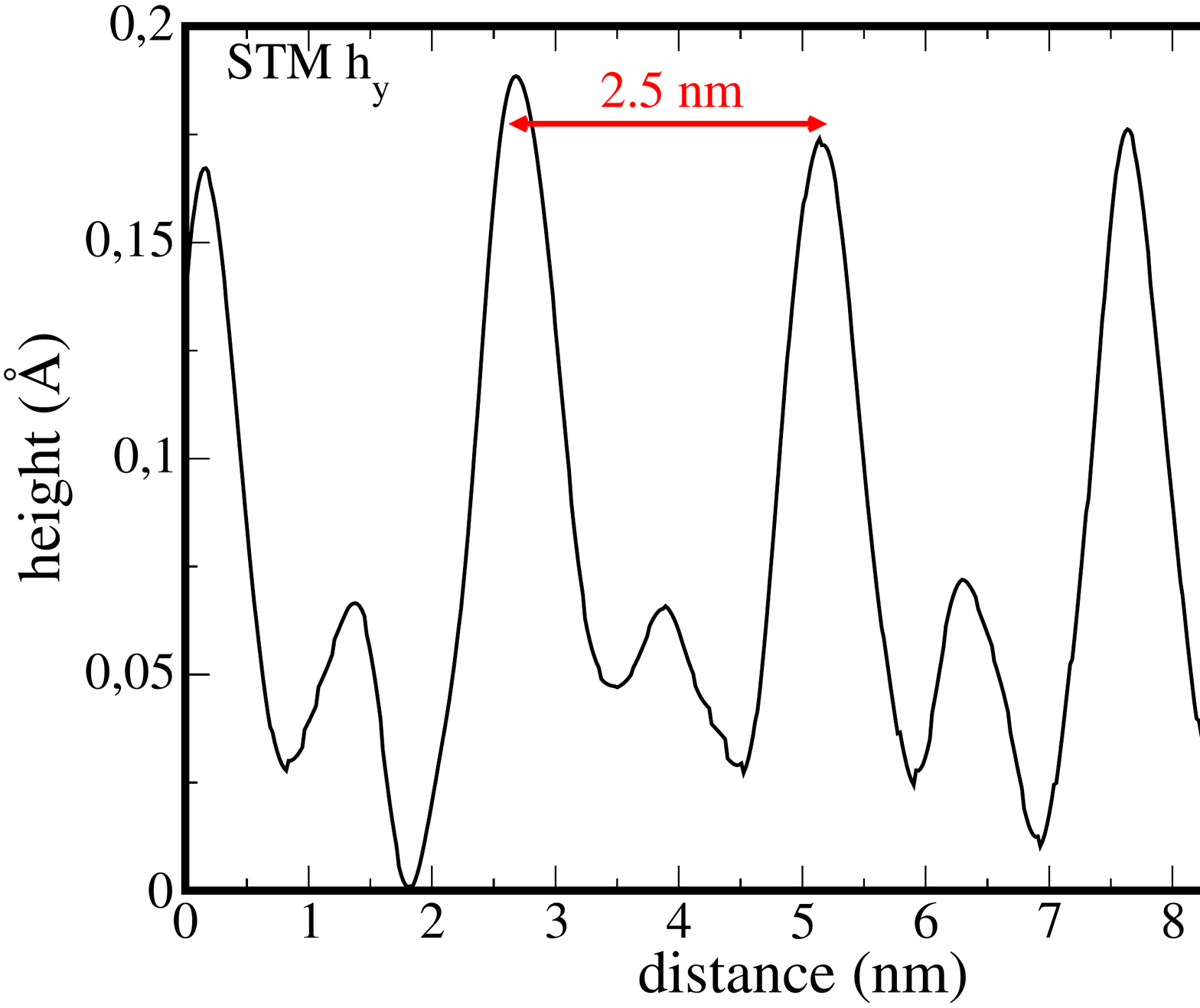}
\caption[]{
Height profiles in the R10 phase along the two high symmetry directions,
cut out within thin sections with the thickness of 
2 $\hbox{\AA}$,
as obtained by CMD (a) and by STM (b) along
the shorter diagonal of the rhomboid supercell
(along the blue line, in Fig 2(d)).
The repeat distances $d_{per}$ are also shown.
(c)-(d): Height profiles along the longer diagonal of the supercell as obtained by: CMD (c)
and by STM (d)
(along the red line in Fig 2(d)).
Note that the secondary peaks between the moir\'{e} peaks can also
be seen.
}
\end{center}
\end{figure*}

\begin{table*}[tb]
\raggedright
\caption[]
{
The summary of various properties obtained for gr/Cu(111)
by classical molecular dynamics simulations using the fitted Abell-Tersoff
potential for the interface.
The main properties of the moir\'{e} superstructures.
}
\begin{ruledtabular}
\begin{tabular}{lcccccccccc}
  & &  &  &  &  &  &  &  &  &   \\
   method & $\Theta$ & $d_{per}$ (nm) & $\xi$ ($\hbox{\AA}$)  & $\xi_{Cu}$ ($\hbox{\AA}$) & $d_{ave}$ ($\hbox{\AA}$) & $a_{gr}$ ($\hbox{\AA}$) & $a_{lm}$ ($\%$) & $E_{adh}$ (eV/C) & $\Delta E$  & $E_{gr}$  \\
  & &  &  &  &  &  &  &  &  &   \\
\hline
  & &  &  &  &  &  &  &  &  &   \\

 CMD & $0.0^{\circ}$ (R0)  & 6.1 & $0.45$ & $0.1$ & 2.99 & 2.440 & 3.56 & -0.146 & 0.0072 & -7.419 \\
  & $2.2^{\circ}$ (R2) & 4.3 & $0.25$ & $0.1$ & 2.99 & 2.440 & 3.56 & -0.146 & 0.04 & -7.384 \\

  & $6.7$ (R7) & 2.5 & $0.14$ & $0.06$ & 3.02 & 2.445 & 3.44 & -0.145 & 0.035 & -7.391 \\

  & $8.7$ (R9) & 1.75 & $0.12$ & $0.05$ & 3.02 & 2.445 & 3.44 & -0.145 & 0.02 & -7.406 \\



 & $10.4^{\circ}$ (R10)  & $1.4$ & $0.1$ & $0.05$ & 2.99 & 2.440 & 3.70 & -0.145 & 0.04 & -7.385 \\
 & $16.1^{\circ}$  (R16) & 0.88 & $0.03$ & $0.02$ & 3.02 & 2.444 & 3.56 & -0.145 & 0.04 & -7.385 \\


  & &  &  &  &  &  &  &  &  &   \\
 EXP & $\sim 0^{\circ}$ (R0) & 6.0$^a$ & $0.35 \pm 0.1^a$ &  n/a &  n/a & 2.46 & 3.53 & -0.11$^b$ & n/a & n/a \\

  & $7^{\circ}$ (R7) & $2.0^c$, 2.2$^d$ & n/a &  n/a &  n/a & n/a & n/a & n/a & n/a & n/a \\

  & $10.4^{\circ}$ (R10), pw$^e$ & 1.5 & $0.15 \pm 0.05$ &  n/a &  n/a & n/a & n/a & n/a & n/a & n/a \\




  & &  &  &  &  &  &  &  &  &   \\
 DFT & $0^{\circ}$  & n/a & n/a & n/a &  3.25$^f$, 3.05$^g$ & n/a & n/a & -0.062$^f$, -0.198$^g$ & n/a & n/a \\
  & &  &  &  &  &  &  &  &  &   \\
\end{tabular}
\footnotetext*[1]{
dimensions:
$\hbox{\AA}$ has been used for corrugation and nm for the periodicity.
pw denotes present work,
$\Theta$ is the rotation angle in degree (the angle between
the line of zig-zag carbon atoms and the adjacent Cu(111) atoms along the line (110) shown in Inset Fig 1(c)).
$d_{per}$ is the periodicity of the minimal moir\'{e} pattern
(the edge length of the rhombus with 4 moir\'{e} humps, $1 \times 1$ supercell),
$\xi$ and $\xi_{Cu}$ are the average corrugation for gr and the topmost
Cu(111) layer ($\hbox{\AA}$).
$d_{ave}$ is the average inter-layer (C-Cu) distance ($\hbox{\AA}$) at the interface.
$a_{gr}$, $a_{lm}$ are the lattice constant of gr ($\hbox{\AA}$) and
the lattice mismatch ($\%$) after simulations ($a_{lm}=100 (a_{s}-a_{gr})/a_{gr}$).
CMD: pw, fitted Abell-Tersoff results with cg minimization with CMD at 300 K.
EXP: the experimental results:
corrugation ($\xi$): our STM results, 
DFT results are also given for comparison \cite{DFT:Ru-Hutter,Batzill,DFT:Ru_Wang}.
All quantities are given per carbon atom.
The adhesion energy $E_{adh}=E_{tot}-E_{no12}$, where $E_{tot}$ is the potential energy/C
after CMD simulation. $E_{no12}$ can be calculated using the final
geometry of CMD simulation with heteronuclear interactions
switched off. Therefore, $E_{adh}$ contains only contributions from
interfacial interactions.
$E_{str,gr}$, the strain energy of the corrugated gr-sheet,
$\Delta E$ (eV/C) is the energy difference with respect to the
perfectly flat periodic gr.
$\Delta E =E_{gr}-E_{gr,flat}$,
where $E_{gr}$ and $E_{gr,flat}=-7.426$ eV/C are the cohesive energy of C atoms
in the corrugated and in the relaxed periodic flat (reference) gr sheet as obtained
by the AIREBO C-potential \cite{airebo}.
$^a$ from refs. \cite{Gao,STM-grcu},
$^b$ from ref. \cite{grcu_adhesion2}, double cantilever beam method:
$E_{adh}$=0.72 J/m$^2$.
$^c$ from ref. \cite{STM-grcu},
$^d$ from ref. \cite{JACS-grcu},
$^e$ present STM work,
$^f$ from ref. \cite{RPA}, obtained by accurate random phase approximation
for a very small modell system.
$^g$ present work: nonlocal vdw-DFT calculation for a small flat system (463 atoms):
hcp: -0.198 eV/C ($d_0=2.95$ $\hbox{\AA}$), hollow: -0.182 eV/C ($d_0=3.09$ $\hbox{\AA}$), vdw-DFT geometry optimized
structures: ontop hcp: -0.350 eV/C, hollow: -0.133 eV/C.
}
\label{T2}
\end{ruledtabular}
\end{table*}

\section{Results and Discussion} 

 The STM images shown in Fig. 1(a) and (b)
reveal an unexpected rotation-misoriented phase with a periodicity of ($1.5 \pm 0.05$) nm that has not been reported so far.
The STM images in Fig. 1(a) and (b) have been acquired at low temperature (78K) using a bias voltage of -230 mV and 5 nA tunneling current, under UHV conditions. From Fig 1(b) the periodicity of the honeycomb lattice has been removed by Fourier filtering, to better visualize the details of the moir\'{e} pattern. The observed periodicity is the smallest graphene superlattice reported on Cu (111) so far.
The new moir\'{e} phase has been observed in several locations on two different samples.

 We found that a larger-angle rotated moir\'{e} superstructure ($\Theta \approx 10.4^{\circ}$)
is able to reproduce our STM findings reported here (see Figs 1(a)-(b)).
The corresponding CMD results can be seen in Figs 1(c)-(d). 
Until now there are only a few experimental works reporting a rotated phase of graphene on Cu(111). One reported rotated phase
is of $\Theta \approx 7^{\circ}$ (R7)
with the periodicity of $d_{per} \approx 2-2.2$ nm \cite{Gao,JACS-grcu}.
Our misoriented phase occurs at considerably larger angle implying
a smaller repeat distance.
In particular, we found ($d_{per} \approx 1.5 \pm 0.05$) nm (STM) and the substantial drop of
corrugation with respect to $\Theta \approx 0^{\circ}$ ($\xi \approx 0.012$ nm).
We denote this phase as R10.
CMD simulations give a slightly smaller $d_{per} \approx 1.4 \pm 0.05$ nm.
This difference can be due to the experimental error. Therefore we propose that a novel stable rotational phase has been identified both theoretically and experimentally.

  The height variation of gr along different directions can be seen
on Figs. 2(a)-2(d). Concerning the larger-angle rotated sample R10 ($\Theta \approx 10.4^{\circ}$),
two main sections are shown along high symmetry directions of the
superlattice. On Figs. 2(a)-(b) the height profiles
are seen along the short diagonal of the minimal rhomboid supercell
from CMD simulations (2(a)) and from STM (2(b)).
Figs. 2(c)-(d) depict the long diagonal height profiles
for CMD simulations (2(c)) and the corresponding STM cross section(2(d)).
The most striking features are the small satellite peaks in between the main moir\'{e} hills, observed in both the measured and simulated cross sections.
 The splitting of the height profiles along distinct 
directions for gr moir\'{e} patterns has also recently been
reported for gr/Ir(111) \cite{Sainio}.
The corrugation for the rotated patterns remains below 0.015 nm
for both directions, while the corrugation of the small
satellite peaks is of just 0.005 nm.

 The observed moir\'{e} superstructure has been reproduced and characterized by CMD simulations (Figs 1(c)-(d))
which required, however, the development of a new interfacial
force field.
Using this method 
we could also identify the two reported phases R0 and R7 phase reported in refs. \cite{Gao,JACS-grcu} by cg-min/CMD simulations.
Moreover, our simulations predict the existence of several other stable rotated phases, such as R2, R9 and R16
which have not been observed yet.
 In particular, 
 the angular dependent Abell-Tersoff potential \cite{tersoff} has been used for the
  interface which was fitted to
nonlocal vdw-DFT potential energy curves of small
modell systems with hcp and hollow binding registries.
No explicit information has been used
in the fitting procedure on corrugation, supercell size or other
experimentally measured quantities to be accounted for \cite{Sule13}.
The new potential together with the AIREBO \cite{airebo}
C-C potential on Cu(111) is capable of reproducing
the moir\'{e} superstructures seen by STM \cite{Gao,STM-grcu} at
nearly zero misorientation angle.
The obtained periodicity ($d_{per}$) of the small supercell is $6.1$
nm in accordance with the experiments \cite{Gao,STM-grcu}.
The resulting corrugation of $\xi \approx 0.4$ $\hbox{\AA}$, is also
in agreement with the available STM data \cite{STM-grcu}. 
  We could also reproduce by CMD simulations the R7 ($\Theta \approx 6.7^{\circ}$) phase with a somewhat
larger repeat distance of $d_{per} \approx 2.5 \pm 0.05$ nm than
found by STM in recent publications ($2.0-2.2$ nm, \cite{Gao,JACS-grcu}).

 In general we report the rapid decrease of the repeat distance with rotation angle, 
for instance for $\Theta=16^{\circ}$ the value of $d_{per} \approx 0.88$ nm
has been found by CMD simulations (see Table II.).
We could not identify stable phases with larger rotation angle with CMD.

 In Table II. the various structural and energetic properties of the simulated rotated structures
have been summarized.
The notable features are the following:
corrugation decreases with $\Theta$ and practically vanishes
for $\Theta > 15^{\circ}$.
The lattice constant of the gr sheet shows no considerable sensitivity
to $\Theta$ and is stabilized at $a_{gr} \approx 2.44$ $\hbox{\AA}$
(the gr lattice is slightly compressed).

 In spite of the significant lattice misfit of $a_{lm} < 3.56$ $\%$
 the aligned gr $\Theta \approx 0^{\circ}$ phase is very close in energy to that of
the perfectly relaxed flat gr, with a vanishingly small energy difference
($\Delta E \approx 0.007$ eV/C) although its corrugation is
is significant ($\xi \approx 0.45$ $\hbox{\AA}$).
This can be attributed to the strain relief in the large coincidence supercells \cite{grru13}.
We found $0.16$ \% lattice mismatch in the minimal supercell
in the R0 phase.
In the case of the misoriented phase reported here with $\Theta=10.4^{\circ}$
the misfit of the coincidence supercells are
$1.47$ \% for the minimal supercell (red rhombus in Figs 1(a) and (b)).

 Concerning the energetic stability with respect to the perfectly
flat periodic gr,
 $\Delta E \approx 0.04$ eV/C (see Table II.,) which is somewhat above
the magnitude of thermal motion ($\sim 0.026$ eV/K at 300 K)
which in principle renders this phase detectable even at room temperature.
Also the energy difference between the aligned and the rotated phase ($\Theta=10.4^{\circ}$) is
in similar range ($\Delta E \approx 0.033$ eV/C).

Interestingly, the large angle rotation ($\Theta > 15^{\circ}$) also does not require much more energy
($\Delta E \approx 0.04-0.05$ eV/C).
This suggests that at elevated temperatures, however, (few hundred K above room temperature)
the complexity of the multiple-oriented rotation domains could be increased
significantly and the number of the large angle rotated regions
could be considerable.
This clearly indicates that the high temperature conditions ($1200$ K) of CVD growth of gr could be sufficient
for the occurrence of a large variety of large angle rotational domains.
The pinning of various rotated domains, once formed during CVD growth
could occur.
Therefore, we predict the coexistence of various misaligned
gr moir\'{e} patterns including ultraflat (highly rotated), low-angle rotated and aligned
regions (corrugated) which are energetically permitted.
It must be noted that
the large angle rotated domains ($\Theta > 15^{\circ}$), do not show moir\'{e} patterns
due to the vanishing corrugation and small repeat distance. 
Similar findings have been reported for gr/Ir(111) \cite{grir_rot}
at much larger rotation angles of $30^{\circ}$ 
while the corrugation was 10 times smaller than for the aligned phase.
Considering that the morphology of gr/Cu(111) is rich even without rotation
\cite{Tapaszto}
it is not easy to directly assign a rotation angle to an observed moir\'{e}
superlattice.
Recent studies suggest similar findings for gr/Pt(111) \cite{rotPt} or
for gr/Ir(111) \cite{rotIr2}.

 The {\em adhesion energy} has also been calculated and is shown
for various rotated superlattices in Table II.
We find a somewhat larger adsorption energy of $-0.145$ eV/C
than by experiment ($-0.11$ eV/C, \cite{grcu_adhesion2}).
In general, the binding energy of gr to Cu(111) is smaller than
in other systems, such as gr/Ru(0001) ($0.17-0.2$ eV/C, see e.g. ref. \cite{Sule13}) and larger
than in the weakest bound systems (e.g. gr/SiO$_2$, $E_{adh} \approx -0.07$ eV/C) \cite{grsio2}.
No considerable dependence of $E_{adh}$ has been found on $\Theta$.


\section{Conclusions}

 We have developed a new interfacial
force field to describe various moir\'{e} superstructures emerging
in graphene on Cu(111) surface.
No such interface interaction potential, adequately describing the moir\'{e} superlattices of graphene/Cu(111) interface has been available so far.
 Using classical molecular dynamics simulations we have shown
that our method can
provide reasonably accurate structural results even for weakly bound
(low-corrugated) extended systems such as graphene on copper.
We also report the experimental observation of a new rotated phase of 1.5 nm periodicity, that has not yet been observed experimentally.
Using 
the new potential we have successfully reproduced the observed novel phase
for rotation angle of 10.4$^{\circ}$, including the fine structure
of the superlattice, without any further parameter adjustment.
Furthermore, 
our model predicts the existence of various superlattices.
Interestingly, while aligned and low-angle rotated structures display a clearly
detectable corrugation, further increasing the misalignment quickly
flattens the graphene sheet. Angles larger than 15$^{\circ}$
are predicted to yield ultra-flat graphene.

%
%

\section{acknowledgement}
LT and CH acknowledges financial support form the Korea Hungary Joint
Laboratory for Nanosciences and OTKA grant K 108753.
The calculations (simulations) have been
done mostly on the supercomputers
of the NIIF center (Hungary).
The availability of codes LAMMPS (S. Plimpton) and OVITO (A. Stukowski) are also greatly acknowledged.



\end{document}